\documentclass[a4paper,11pt]{article}

\usepackage{contribution}

\newcommand{\weblink}[2][]{%
    \ifthenelse{\equal{#1}{}}%
    {\textnormal{\url{#2}}}%
    {\textnormal{\href{#2}{#1}}}%
}

\def\beq{\begin{equation}}
\def\eeq#1{\label{#1}\end{equation}}
\def\eeqn{\end{equation}}

\def\beqa{\begin{eqnarray}}
\def\eeqa#1{\label{#1}\end{eqnarray}}
\def\eeqan{\end{eqnarray}}

\let\bar=\overbar

\def\Dslash{\not{\hbox{\kern-4pt $D$}}}
\def\dslash{\not{\hbox{\kern-2pt $\del$}}}

\def\msb{{\bar{\ssstyle M \kern -1pt S}}}

\newcommand{\contribution}[7][]{%
  \clearpage
  \thispagestyle{plain}
  \ifthenelse{\equal{#1}{}}
  {\hypersetup{pdftitle={#2}}}
  {\hypersetup{pdftitle={#1}}}
  \hypersetup{pdfauthor={{#3} {#4}}}
  {\centering\normalfont\LARGE\bfseries\sffamily #2 \par\nobreak}
  \lhead{}
  \chead{%
    \textit{\footnotesize XIV International Conference on Hadron Spectroscopy
      (\weblink[\textit{hadron2011}]{http://www.hadron2011.de}), 13-17 June 2011, Munich, Germany}%
  }
  \rhead{}
  \bigskip
  \begin{center}
    {#3} {#4}\ifthenelse{\equal{#6}{}}{}{\footnote{\weblink[#6]{mailto:#6}}}
    \ifthenelse{\equal{#7}{}}{}{#7} \\
    \textit{#5}
  \end{center}
  \bigskip
}

\renewcommand{\abstract}[1]{%
  \begin{center}
    \begin{minipage}{0.85\textwidth}
      \begin{footnotesize}
        #1
      \end{footnotesize}
    \end{minipage}
  \end{center}
  \bigskip
}

\begin{document}

%
%
%
%
%
{  


%

\contribution
{Hadronization in Nuclei -- Multidimensional Study}  
{Inti}{Lehmann}  
{Department of Physics and Astronomy \\
  University of Glasgow \\
  Glasgow, G12 8QQ, Scotland/UK
}  
{inti.lehmann@glasgow.ac.uk}  
{on behalf of the HERMES Collaboration}  
%

\abstract{%
  Hadron multiplicities in semi-inclusive deep-inelastic
  scattering were measured on neon, krypton and xenon targets relative
  to deuterium at an electron-beam energy of 27.6\,GeV at HERMES.
  These ratios were determined as a function of the virtual-photon
  energy $\nu$, its virtuality $Q^2$, the fractional hadron energy $z$ and
  the transverse hadron momentum with respect to the
  virtual-photon direction $p_t$.  Dependences were analysed separately for
  positively and negatively charged pions and kaons as well as protons
  and antiprotons in a two-dimensional representation.  These results
  will help to constrain mechanisms and models of hadronization much
  more decisively than by the use of integrated results as
  traditionally done.  A few features particular to the two-dimensional representation are
  highlighted in this contribution. 
}
%


Semi-inclusive production of hadrons in deep-inelastic lepton nucleus
scattering (SIDIS) provides a way to study quark fragmentation or
hadronization. Lepto-production of hadrons has
the virtue that the energy and momentum transfered to the hit parton
are well determined, as it is ``tagged'' by the scattered lepton.  In
these studies the nucleus is basically used as a scale probe of the
underlying hadronization mechanism: by using nuclei of increasing size
one can investigate the the space(time) development of hadronization.

The ratio of normalised yields $Y_A^h$ on neon (Ne), krypton (Kr) and
xenon (Xe) targets, denoted by $A$, compared to the same quantity on a
deuterium $D$ target: $R_A^h = Y_\mathrm{A}^h / Y_\mathrm{D}^h$
was measured with the HERMES
spectrometer~\cite{herm0} at DESY, where $h$ indicates positively and negatively charged
pions ($\pi^{+/-}$) and kaons (K$^{+/-}$), protons (p) and antiprotons
($\bar{\mathrm{p}}$).  For the first time a two-dimensional
representation is chosen for all hadrons separately.  This allows to
observe features that are hidden when integrating over large kinematic
ranges. Details, references to previous measurements and theoretical
studies can be found in Ref.~\cite{publication}.

The results show, for example (see Fig.~\ref{f:nu_z}), that $\pi^+$
and $\pi^-$ behave similarly. However, their dependences with the
virtual-photon energy $\nu$ change with the fraction carried by
the hadron $z$.  $K^+$ show different features compared to $K^-$
which could be due to their different quark content.  Particularly
striking is the behaviour of protons, which show completely different
trends in different ranges of $z$. Presumably, this is due to
a sizable contribution of final-state interactions, such as
knock-out processes, in addition to the fragmentation process.

\begin{figure}[htb]
\includegraphics[width=.48\textwidth]{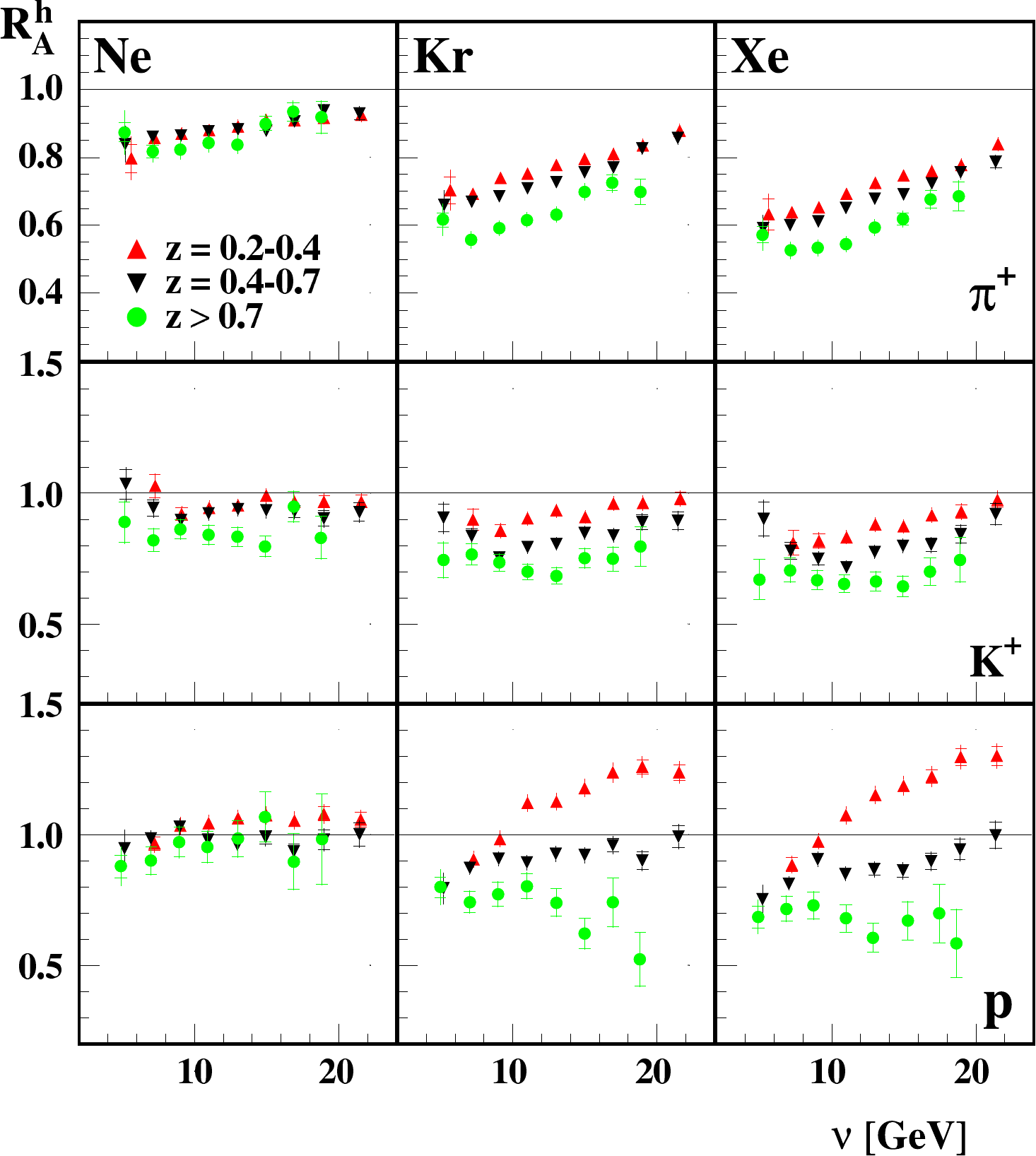}\hfill{ }
\includegraphics[width=.48\textwidth]{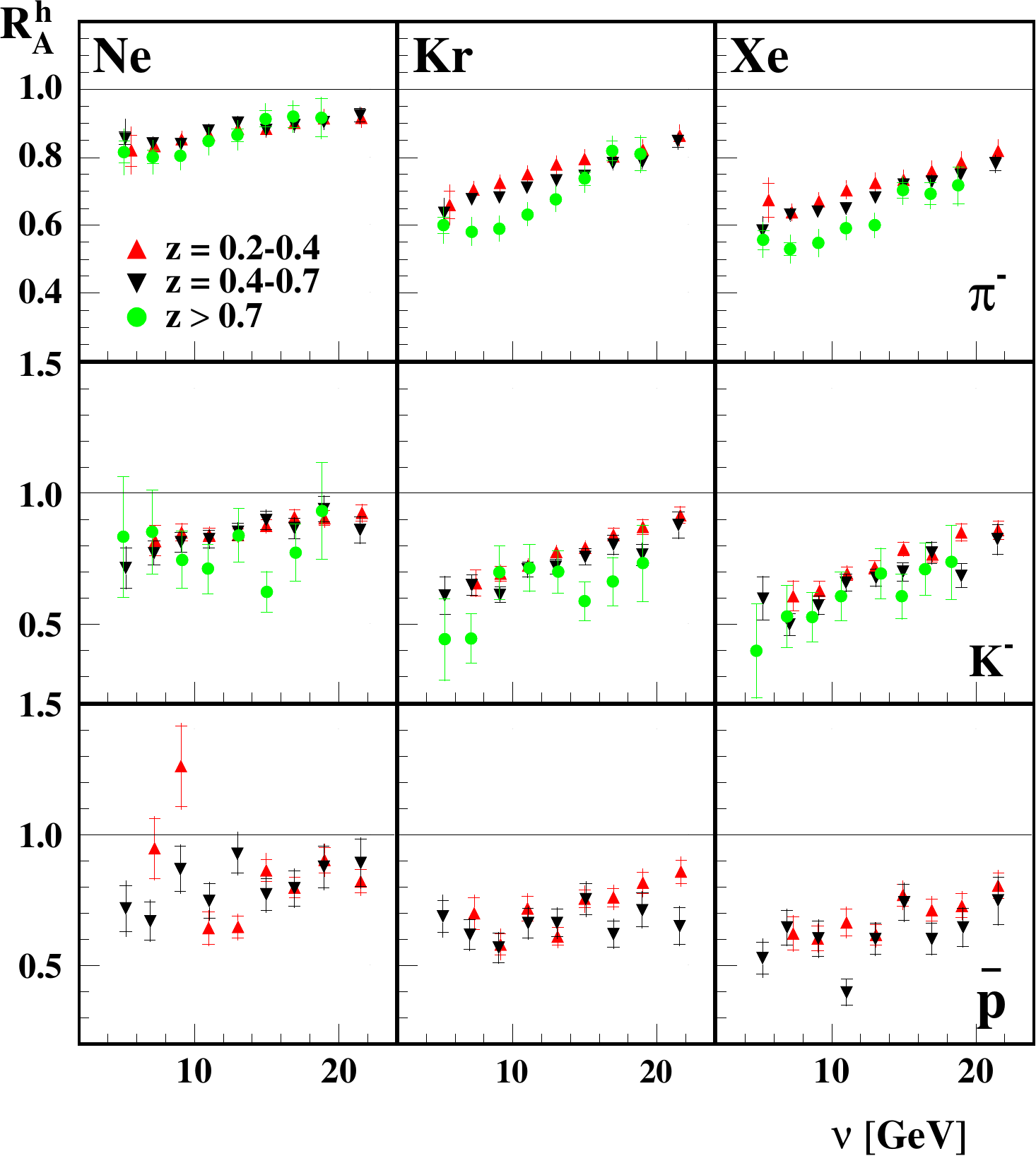}
\caption{Dependence of $R_{A}^h$ on $\nu$ for positively and
  negatively charged hadrons for three slices in $z$ as indicated
  in the legend.  The inner and outer error bars indicate the
  statistical and total uncertainties, respectively.  For the latter
  the statistical and systematic bin-to-bin uncertainties were added
  in quadrature.  In addition, scale uncertainties of 3\%, 5\%, 4\%, and
  10\% are to be considered for pions, kaons, protons and
  antiprotons, respectively.}
\label{f:nu_z}
\end{figure}

In conclusion, the two-dimensional distributions of $R_A^{h}$ for
identified $\pi^+$, $\pi^-$, $\mathrm{K}^+$, $\mathrm{K}^-$,
protons and antiprotons, measured at HERMES~\cite{publication}, provide detailed
information which is generally not accessible in the one-dimensional
distributions (in which all kinematic variables except one are
integrated over, as has been traditionally done).  These new detailed
data are expected to be an essential ingredient for constraining
models of hadronization and, hence, improving our understanding of
hadron formation.

We gratefully acknowledge the DESY management for its support, the
staff at DESY and the collaborating institutions for their significant
effort, and our national funding agencies for financial support.

%

}  


\end{document}